\pdfoutput=1
\documentclass[conference]{IEEEtran}
\IEEEoverridecommandlockouts
\usepackage[utf8]{inputenc}
\usepackage[T1]{fontenc}

\usepackage{cite}
\usepackage{amsmath,amssymb,amsfonts}

\usepackage{graphicx}
\usepackage{xcolor}
\usepackage{tikz}
\usepackage{pgfplots}
\pgfplotsset{compat=1.18}

\usepackage{booktabs}
\usepackage{array}
\usepackage{multirow}
\usepackage{longtable}
\usepackage{rotating}

\usepackage{float}
\usepackage{subcaption}

\usepackage{algorithm}
\usepackage{algpseudocode}
\usepackage{listings}

\usepackage[left=54pt,right=37pt,top=105pt,bottom=54pt]{geometry}
\usepackage{changepage}
\usepackage{textcomp}

\usepackage{hyperref}

\lstdefinelanguage{json}{
    basicstyle=\ttfamily\footnotesize,
    breaklines=true,
    showstringspaces=false,
    string=[s]{"}{"},
    stringstyle=\color{red},
    comment=[l]{//},
    morecomment=[s]{/*}{*/},
    commentstyle=\color{gray}\ttfamily,
    keywordstyle=\color{blue}\bfseries
}

\newgeometry{left=54pt,right=37pt,top=122pt,bottom=54pt}

\begin{document}


\title{HumAIne-Chatbot: Real-Time Personalized Conversational AI via Reinforcement Learning \\
}

\author
{
\IEEEauthorblockN{1\textsuperscript{st} Georgios Makridis}
\IEEEauthorblockA{\textit{Department of Digital Systems} \\
\textit{University of Piraeus}\\
Piraeus, Greece \\
gmakridis@unipi.gr}
\and
\IEEEauthorblockN{2\textsuperscript{nd} George Fragiadakis}
\IEEEauthorblockA{\textit{Department of Informatics and Telematics} \\
\textit{Harokopio University of Athens}\\
Athens, Greece \\
gfragi@hua.gr}
\and
\IEEEauthorblockN{3\textsuperscript{rd} Jorge Oliveira }
\IEEEauthorblockA{\textit{HEI-Lab} \\
\textit{University Lusofona/Immersive Lives}\\
Lisbon, Portugal\\
jorge.oliveira@immersivelives.pt}
\and
\IEEEauthorblockN{4\textsuperscript{th} Tomaz Saraiva}
\IEEEauthorblockA{\textit{Immersive Lives} \\
\textit{Immersive Lives}\\
Lisbon, Portugal \\
tomaz.saraiva@immersivelives.pt}
\and
\IEEEauthorblockN{5\textsuperscript{th} Philip Mavrepis}
\IEEEauthorblockA{\textit{Department of Digital Systems} \\
\textit{University of Piraeus}\\
Piraeus, Greece \\
pmav@unipi.gr}
\and
\IEEEauthorblockN{6\textsuperscript{th} Georgios Fatouros}
\IEEEauthorblockA{\textit{Department of Digital Systems} \\
\textit{University of Piraeus}\\
Piraeus, Greece \\
gfatouros@unipi.gr}
\and
\IEEEauthorblockN{7\textsuperscript{th} Dimosthenis Kyriazis}
\IEEEauthorblockA{\textit{Department of Digital Systems} \\
\textit{University of Piraeus}\\
Piraeus, Greece \\
dimos@unipi.gr}
}

\maketitle

\begin{abstract}
Current conversational AI systems often provide generic, one-size-fits-all interactions that overlook individual user characteristics and lack adaptive dialogue management. To address this gap, we introduce \textbf{HumAIne-chatbot}, an AI-driven conversational agent that personalizes responses through a novel user profiling framework. The system is pre-trained on a diverse set of GPT-generated virtual personas to establish a broad prior over user types. During live interactions, an online reinforcement learning agent refines per-user models by combining implicit signals (e.g. typing speed, sentiment, engagement duration) with explicit feedback (e.g., likes and dislikes). This profile dynamically informs the chatbot dialogue policy, enabling real-time adaptation of both content and style.
To evaluate the system, we performed controlled experiments with 50 synthetic personas in multiple conversation domains. The results showed consistent improvements in user satisfaction, personalization accuracy, and task achievement when personalization features were enabled. Statistical analysis confirmed significant differences between personalized and nonpersonalized conditions, with large effect sizes across key metrics. These findings highlight the effectiveness of AI-driven user profiling and provide a strong foundation for future real-world validation. 
\end{abstract}






\section{Introduction}
\label{introduction}
The Information Age, marked by the proliferation of advanced computational systems and significant data generation, has guided in a transition from Industry 4.0 to Industry 5.0 \cite{xu2021industry}. Industry 5.0 emphasizes collaboration between humans and robots, incorporating resilience and sustainability goals into industrial practices. While Industry 4.0 focused on technologies such as the Internet of Things (IoT) and big data - with Artificial Intelligence (AI) as its driving force, facilitating the development of innovative tools and methods \cite{makridis2020predictive}, \cite{makridis2023deep}, \cite{kotios2022personalized} - Industry 5.0 reintroduces human, environmental, and social considerations into technological advancements. Alongside this shift, the rise of large language models (LLMs) and Generative AI (GenAI) has further enhanced human-machine collaboration.

This evolution has significantly influenced the domain of conversational AI. In recent years, chatbots have become increasingly prevalent across various sectors, transforming how businesses interact with customers and optimizing internal operations. The advent of LLMs like GPT-3 and GPT-4 \cite{brown2020language} has enabled chatbots to generate coherent and contextually rich dialogue across diverse topics, expanding the boundaries of conversational AI.

Despite their widespread adoption, current chatbots face significant challenges in delivering a personalized and efficient interaction experience. Two primary issues are particularly notable:
\begin{itemize}
\item 	\textbf{Lack of Personalization}: Many chatbots today offer generic responses that do not consider individual user characteristics, preferences, or expertise \cite{liu2023implicit}. 
\item \textbf{Lack of Personalization metrics}: Despite their increasing use, evaluating chatbots remains a challenging research problem, with no universally accepted metric for effectiveness, particularly from user-centric perspectives such as motivation and satisfaction \cite{radziwill2017evaluating}, \cite{io2017chatbots}
\end{itemize}

\par This paper tackles the challenge of personalized conversational AI by introducing \textit{HumAIne-chatbot}, an AI-driven chatbot designed to enhance human-machine interactions through adaptive user profiling. The main contributions of this paper are:
\begin{enumerate}
    \item Developing a user profiling mechanism that integrates implicit and explicit metrics to personalize chatbot responses according to individual user characteristics, preferences, and expertise. 
    \item Leveraging an innovative virtual personas mechanism to enhance user profiling consistency and robustness.
    \item Offering a mechanism for transforming user profiling outputs into actionable prompt refinement, enabling more effective user personalization.
\end{enumerate}

In this context, \textit{HumAIne-chatbot} leverages AI-driven user profiling to enhance user experience and effectiveness through two personalization methods. First, it enriches prompts to tailor responses according to the user's knowledge level and expertise, ensuring that the interaction is adaptive to individual user needs. 

To achieve these personalization goals, \textit{HumAIne-chatbot} employs a combination of implicit and explicit metrics to \"know\" the user through AI-based profiling. Implicit metrics include session duration, response time, sentiment analysis, frequency of grammatical mistakes, language complexity, and typing speed. Explicit metrics encompass direct user feedback, profile completeness, and survey-based satisfaction ratings \cite{xu2022adaptive, liu2023implicit}. By utilizing these collected metrics, the AI-based user profiling allows for a personalized interaction experience. Similar approaches have been used in previous studies to gather detailed measurements of chatbot interactions, such as \cite{CercasCurry2020, Deriu2021, Fadhil2018}.

The architecture of \textit{HumAIne-chatbot} integrates several key components, including an Interaction Tracking Engine, AI-Driven User Profiler, Prompt Manager, Dialogue History and Analysis Module, and Dialogue Recommendation Module. This comprehensive system enables the chatbot offer personalized interactions based on the individual user's behavior and context. By leveraging personalized prompt enrichment, \textit{HumAIne-chatbot} ensures that its responses are both relevant and contextually adapted to meet each user's unique needs and capabilities.

Regarding the novel user profiling mechanism that created and utilizes  virtual personas for LLMs by aggregating detailed backstories as described in \cite{moon2024virtual}. This approach facilitates the training process of the AI user profiling approach, allowing the chatbot to generate personalized responses. By incorporating these virtual personas, \textit{HumAIne-chatbot} can better align its conversational strategies with the broader context of the user's values, beliefs, and objectives.

The remainder of the paper is structured as follows: Section 2 provides an overview of current conversational AI approaches, particularly focusing on personalization techniques and user profiling. Section 3 introduces the architecture of \textit{HumAIne-chatbot}, detailing its components and the rationale behind integrating advanced user profiling mechanisms. Section 4 presents the plan for a user study, and Section 5 concludes the paper.

\section{Related Work}

The development of human-centric chatbots is rooted in decades of research in AI, natural language processing (NLP), and human-computer interaction. This section provides an overview of the existing work that informs the design of \textit{HumAIne-chatbot}, focusing on chatbot evolution, personalization techniques, and adaptive conversation management.

\subsection{Chatbot Evolution}

The evolution of chatbots can be traced back to early systems such as \textit{ELIZA}, developed in the 1960s, which employed basic pattern matching to simulate conversation \cite{weizenbaum1966eliza}. 

With advancements in NLP, more sophisticated chatbots powered by LLMs, such as GPT-3 and GPT-4, have been introduced \cite{brown2020language}. These LLMs enable the generation of coherent and contextually rich dialogues across various domains. However, despite these capabilities, LLM-based chatbots often provide generic responses that are not well adapted to the specific needs of individual users. ChatGPT, for example, is recognized as a leader in the field due to its public API, extensive training data, and adaptability across diverse tasks \cite{jasper23report}. It has seen widespread adoption in sectors such as healthcare, education and finance \cite{sallam2023chatgpt}, \cite{makridis2024fairylandai} \cite{fatouros2023transforming} Moreover, MarketSense-AI is a notable application of GPT-4 in the financial domain, leveraging the Chain-of-Thought (CoT) approach to effectively elucidate investment decisions \cite{fatouros2024can}.

Another noteworthy application of LLMs, particularly the GPT-4 model, is presented by \cite{mavrepis2024xai}, which introduces a human-centric XAI tool named "x-[plAIn]". This tool represents a novel approach in Explainable Artificial Intelligence (XAI), aimed at making XAI more accessible to non-experts. It utilizes a custom LLM to generate clear, audience-specific summaries of XAI methods tailored to different groups. By adapting the explanations based on the audience's knowledge level and interests, this model improves decision-making and accessibility.

\textit{HumAIne-chatbot} is designed to be flexible, allowing it to integrate with any API-based LLM, providing adaptability and scalability in leveraging different models. This capability ensures that \textit{HumAIne-chatbot} can incorporate the most suitable LLMs for specific use cases, thereby enhancing personalization and conversation quality.

\subsection{Personalization in Chatbots}

Personalization is an important feature to improve user satisfaction in conversational systems. Existing frameworks, such as the user-adaptive model by \cite{xu2022adaptive}, use explicit user input to tailor chatbot responses, requiring users to provide preferences and other personal data. This explicit personalization approach, while effective to some extent, presents privacy challenges and requires ongoing user effort.
Most chatbot evaluation research has focused on technical metrics such as passing the Turing test, while there is a significant gap in studies that address user-centric aspects such as user satisfaction and motivation \cite{io2017chatbots} \cite{brandtzaeg2017people}. Implicit personalization, which involves deriving user preferences from interaction behaviors, has been proposed to overcome these issues. \cite{liu2023implicit} explored metrics like response time and typing speed to build user profiles, enabling adaptive conversation strategies. \cite{serban2018survey} emphasized the importance of holistic benchmarking frameworks that assess the quality, coherence, and adaptability of user interaction. However, these approaches still face limitations, particularly in effectively anticipating user needs or adjusting response depth in real-time for more nuanced and complex contexts.

Adaptive conversation management aims to make chatbot interactions more engaging and contextually relevant. Traditional scripted approaches often limit the chatbot's ability to handle dynamic conversation paths. Deep reinforcement learning (RL) has been introduced to enable chatbots to learn optimal interaction policies dynamically \cite{li2016deep}. 

\textit{HumAIne-chatbot} goes beyond the state of the art by utilizing both explicit and implicit personalization methods. Additionally, we introduce an innovative approach for training the user-profiler based on \cite{moon2024virtual}, which creates detailed virtual personas from user backstories. 

\subsection{Chatbots in Specific Domains}

Chatbots are increasingly being utilized across a wide range of fields, from customer service to education, healthcare, and entertainment, serving both personal and professional purposes \cite{venkatesh2018evaluating}. In customer service, chatbots like \textit{Dialogflow} by Google and \textit{ChatGPT} handle frequent customer inquiries and provide automated support. In healthcare, chatbots such as \textit{Woebot} offer personalized support for mental health \cite{fitzpatrick2017woebot}, demonstrating the value of tailored interaction. 

In the restaurant industry, chatbots are being employed to enhance customer service processes and provide personalized recommendations \cite{romero2023approach}. These AI-powered assistants use advanced techniques such as embedding and context adaptation to generate accurate and relevant responses to customer queries.
In the realm of elderly care, companion bots are gaining popularity as a means to combat loneliness and provide assistance with daily tasks \cite{lappromrattana2023quick}. These bots combine features of mobile robots and chatbots, offering functionalities such as medication reminders and entertainment options, thereby enhancing the quality of life for elderly individuals.
The export business sector has also seen the integration of conversational agents to accelerate business processes \cite{jamil2023systematic}. These AI-powered chatbots use natural language processing and machine learning to facilitate real-time conversations, providing support in areas such as marketing and customer assistance.
The banking sector has also adopted AI chatbots to improve customer loyalty \cite{alghiffari2023antecedents}. Factors such as AI chatbot service quality, cognitive trust, and customer satisfaction play crucial roles in determining the effectiveness of these chatbots in banking applications.

\section{Proposed Methodology}
In this section, we present our framework for personalizing chatbot interactions to improve user satisfaction. Our approach involves measuring various metrics that serve two main purposes: inputting data into the user profiler to personalize the HumAIne chatbot and evaluating the overall performance of the chatbot. The metrics used for personalization, as outlined in Table~\ref{tab:personalization}, include both implicit and explicit measurements collected during interactions. Implicit metrics like session duration, response time, typing speed, grammatical accuracy, language complexity, and affect provide insights into the user's behavior and communication style without requiring direct input. Explicit metrics, such as responses to pre-session questions, offer direct information from the user to inform the personalization process.

Metrics used to evaluate the overall performance of the chatbot, as shown in Table~\ref{tab:evaluation}, focus on assessing user satisfaction and the quality of the interaction. These include explicit metrics like user feedback ratings, comments, usability assessments, and measures of satisfaction such as expectation, impression, navigability, and engagement.

By combining these metrics, we aim to create a framework that not only personalizes the chatbot experience but also continually assesses and improves the chatbot's performance.

\begin{table*}[htbp]
\centering
\caption{Metrics for Personalization (Inputs to User Profiler)}
\begin{tabular}{|p{2cm}|p{3cm}|p{6cm}|p{2cm}|p{2cm}|}
\hline
\textbf{Perspective} & \textbf{Category} & \textbf{Metrics} & \textbf{Type} & \textbf{Measurement} \\
\hline
General Interaction & Session Duration & Length of Interaction Time & Quantitative & Implicit \\
\cline{2-5}
& Response Time & Average Time to Respond & Quantitative & Implicit \\
\cline{2-5}
& Typing Speed & Characters per Second & Quantitative & Implicit \\
\hline
Linguistic & Grammatical Accuracy & Frequency of Grammatical Errors & Quantitative & Implicit \\
\cline{2-5}
& Complexity of Language & Avg. Sentence Length, Type-Token Ratio & Quantitative & Implicit \\
\hline
User Experience & Affect & Emotional Information, Personality Traits, Engagement & Qualitative, Quantitative & Implicit \\
\hline
General Interaction & User Elicitation & Responses to Pre-Session Questions & Qualitative & Explicit \\
\hline
\end{tabular}
\label{tab:personalization}
\end{table*}

\begin{table*}[]
\centering
\caption{Metrics for Evaluating Overall Performance of the Chatbot}
\begin{tabular}{|p{2cm}|p{3cm}|p{6cm}|p{2cm}|p{2cm}|}
\hline
\textbf{Perspective} & \textbf{Category} & \textbf{Metrics} & \textbf{Type} & \textbf{Measurement} \\
\hline
General Interaction & User Feedback & Ratings, Comments & Qualitative & Explicit \\
\hline
User Experience & Usability & Task Completion, Getting Assistance & Qualitative, Quantitative & Explicit \\
\cline{2-5}
& Satisfaction & Expectation, Impression, Navigability, Engagement & Qualitative, Quantitative & Explicit \\
\hline
\end{tabular}
\label{tab:evaluation}
\end{table*}

\subsection{Implicit Metrics}

\subsubsection{Session Duration}

Session Duration ($SD$) measures the length of time a user interacts with the chatbot in a single session. It can be calculated as:

\begin{equation}
SD = t_{\text{end}} - t_{\text{start}}
\end{equation}

where $t_{\text{start}}$ is the timestamp when the session begins, and $t_{\text{end}}$ is when the session ends.

\subsubsection{Response Time}

Response Time ($RT$) tracks how quickly the user responds to the chatbot's prompts. For each turn $i$, the response time is:

\begin{equation}
RT_i = t_{\text{user\_response}}^i - t_{\text{chatbot\_prompt}}^i
\end{equation}

The average response time over $N$ interactions is:

\begin{equation}
\overline{RT} = \frac{1}{N} \sum_{i=1}^{N} RT_i
\end{equation}

\subsubsection{Sentiment Analysis}

Sentiment Analysis involves quantifying the emotional tone of user responses. A sentiment score ($SS$) can be assigned to each user message using a sentiment analysis function $S(\cdot)$:

\begin{equation}
SS_i = S(\text{User Message}_i)
\end{equation}

The average sentiment score over $N$ messages is:

\begin{equation}
\overline{SS} = \frac{1}{N} \sum_{i=1}^{N} SS_i
\end{equation}

\cite{Liu2012}

\subsubsection{Grammatical Mistakes Frequency}

Grammatical Mistakes Frequency ($GMF$) tracks the number of grammatical errors per word in the user's responses. For message $i$:

\begin{equation}
GMF_i = \frac{E_i}{W_i}
\end{equation}

where $E_i$ is the number of grammatical errors and $W_i$ is the total number of words in message $i$. The average over $N$ messages is:

\begin{equation}
\overline{GMF} = \frac{1}{N} \sum_{i=1}^{N} GMF_i
\end{equation}

\subsubsection{Complexity of Language}

The Complexity of Language ($CL$) can be quantified using metrics like average sentence length ($ASL$) and type-token ratio ($TTR$):

\begin{equation}
CL_i = \alpha \times ASL_i + \beta \times TTR_i
\end{equation}

where $\alpha$ and $\beta$ are weighting factors, $ASL_i$ is the average number of words per sentence, and $TTR_i = \frac{\text{Number of Unique Words}}{\text{Total Words}}$ in message $i$.

\subsubsection{Typing/Interaction Speed}

Typing Speed ($TS$) measures how quickly the user types their responses:

\begin{equation}
TS_i = \frac{C_i}{t_{\text{user\_response}}^i - t_{\text{user\_start}}^i}
\end{equation}

where $C_i$ is the number of characters in message $i$, $t_{\text{user\_start}}^i$ is when the user starts typing, and $t_{\text{user\_response}}^i$ is when they send the message.

\subsection{Explicit Metrics}

Explicit User Feedback can be quantified using a feedback score ($FS$) based on "likes" of the responses:

\begin{equation}
FS = \frac{\sum_{i=1}^{n} L_i}{n}
\end{equation}

where $L_i$ represents the number of likes for response $i$, and $n$ is the total number of responses considered.

\subsubsection{Session-Based User Elicitation}

To enhance user profiling before each session, we propose a novel approach where users are asked 2-3 targeted questions before starting a conversation. These questions are designed to gather updated user preferences, interests, or context-specific information that can enhance personalization during the interaction. This approach serves as an explicit metric for understanding the user's needs at the beginning of each session and tailoring responses accordingly.

Asking these targeted questions before each session raises the issue of user elicitation, which involves finding the balance between asking relevant questions and not overwhelming the user. However, this elicitation process is not a straightforward task. Quantifying the performance of interview chatbots is a significant challenge. Developing a computational framework to measure the effectiveness of these chatbots is crucial for their improvement \cite{han2021designing}.

\subsubsection{Survey-Based Satisfaction}

Survey-Based Satisfaction ($SBS$) aggregates responses from post-session surveys:

\begin{equation}
SBS = \frac{1}{M} \sum_{j=1}^{M} R_j
\end{equation}

where $R_j$ is the user's response to question $j$ on a standardized scale, and $M$ is the number of survey questions.

\subsection{Proposed Personalized Chatbot Architecture}

\begin{figure*}[h!]
\centering
\includegraphics[width=0.8\textwidth]{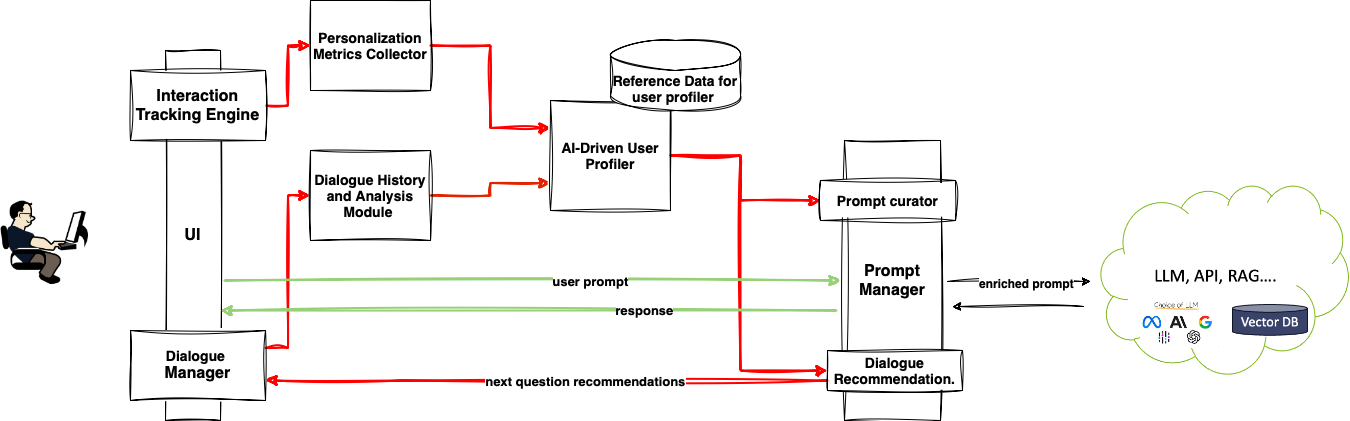}
\caption{Architecture of the \textit{HumAIne-chatbot} system, illustrating the flow of data between components. }
\label{fig:trajectory}
\end{figure*}

\begin{figure*}[h!]
\centering
\includegraphics[width=0.8\textwidth]{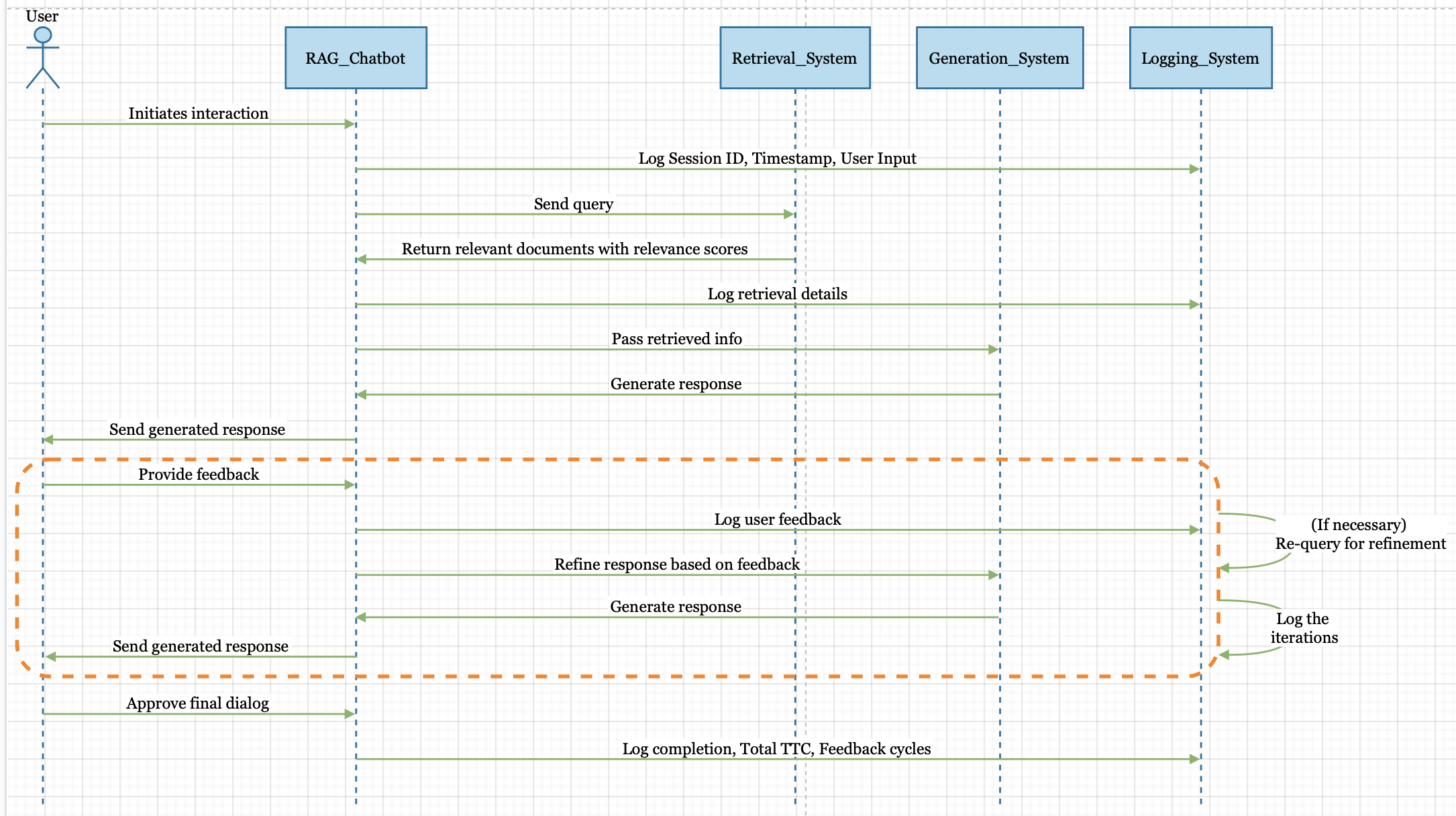}
\caption{Sequence Diagram of Interaction Flow of \textit{HumAIne-chatbot} based on a RAG approach} \label{fig:sequence}
\end{figure*}

This section provides a detailed description of each component within the personalized chatbot system depicted in Figure~\ref{fig:trajectory}. The components collectively enable the personalized and adaptive conversation flow, using advanced user profiling and recommendation strategies. While the sequence diagram \ref{fig:sequence} illustrates the interaction between the user and a RAG chatbot system, showcasing how the system retrieves relevant information, generates responses, and incorporates user feedback in an iterative process until the final response is approved.

\subsubsection{User Interface (UI) with Integrated Dialogue Management and Tracking}

The \textbf{User Interface UI)} is the central point of interaction between the user and the chatbot system. It manages user input, displays responses, and handles interaction feedback, providing a seamless and engaging experience that ensures the conversation flow remains intuitive and accessible. The UI integrates both the Dialogue Manager and the Interaction Tracking Engine, allowing for cohesive management of the conversation and real-time adaptation based on user behavior.

The \textbf{Dialogue Manager} within the UI is responsible for controlling the overall flow of the conversation. It coordinates with other modules, such as the Dialogue History Module, AI-Driven User Profiler, and Prompt Manager, to maintain a coherent and adaptive conversational flow. By managing turn-taking, topic transitions, and conversation pacing, the Dialogue Manager ensures that the chatbot maintains a natural dialogue structure. It also integrates recommended prompts and manages user expectations, thereby enhancing the user experience.

The \textbf{Interaction Tracking Engine} embedded in the UI captures user activity during conversations. It continuously monitors all user interactions, including input characteristics such as response time, session duration and user behavior patterns. These metrics are then forwarded to other components for further analysis and personalization purposes.

\subsubsection{Personalization Metrics Collector}
The \textbf{Personalization Metrics Collector} is responsible for gathering and analyzing metrics during user interactions, such as response times, engagement level, and other behavioral metrics. These metrics are critical for generating detailed user profiles, enabling adaptive personalization. The collected data feeds into the AI-Driven User Profiler, which refines user profiles for more precise personalization.

\subsubsection{Dialogue History and Analysis Module}
The \textbf{Dialogue History and Analysis Module} is responsible for extracting and analyzing text-based metrics, including sentiment, grammatical accuracy, and complexity of language. This module also manages historical conversation data, including past user responses, previously discussed topics, and interaction patterns. The text-based metrics gathered by this module are forwarded to the AI-Driven User Profiler to enhance user profiling. Additionally, it plays a vital role in maintaining conversational continuity, allowing the chatbot to seamlessly reference past interactions and sustain context across the session. 

\subsubsection{AI-Driven User Profiler with Reference Data Integration}

The \textbf{AI-Driven User Profiler} is responsible for creating comprehensive user profiles using collected metrics, historical data, and reference data. It leverages ML models to analyze user interactions and extract information such as interests, and communication preferences. The integration of reference data provides a knowledge base against which user behavior and interactions are benchmarked. This reference data includes general user profiles created with the virtual personas for LLMs.

Initially, the AI model for user profiling will be trained as a supervised ML model using the gathered metrics and reference data. The supervised model will help identify relationships between user behavior metrics and corresponding user traits, forming the basis of user profiles. As the system matures, we will transition to a reinforcement learning (RL) model to further enhance personalization capabilities. The RL model will allow for continuous learning through interaction, optimizing the chatbot's responses based on rewards derived from user engagement and satisfaction metrics.

To effectively utilize advanced techniques in personalized chatbots and create virtual personas for simulating chatbot interactions, we will incorporate the proposed metrics into our framework. This approach enables us to train an AI model for user profiling based on the gathered metrics from simulated behavior.

\paragraph{Phase I: Virtual Persona Pre‑training} 
We leverage the \emph{Anthology} of GPT‑generated backstories to simulate a diverse set of virtual personas. Each backstory encodes rich demographic and behavioral traits (age, occupation, language style), conditioning an LLM to 
"play" a consistent user archetype~\cite{Moon2024VirtualPersonas}. By engaging these personas in multi‑turn dialogues, we collect a large synthetic corpus with "ground‑truth" profile labels. The profiler is first trained in a supervised manner on this corpus, learning to infer latent user attributes and preferences from conversational cues.  

This approach aligns with the VirtualXAI framework, which uses GPT‑generated personas for qualitative assessment of explainability methods~\cite{Makridis2024VirtualXAI}.

\paragraph{Phase II: Online Reinforcement Learning Adaptation} 
In live deployment, each real user interaction constitutes an RL episode. The profiler agent receives a reward $r_t$ derived from engagement signals: conversational length, return sessions, and explicit feedback (likes, sentiment)~\cite{Li2016DeepRL,Wang2024UserSimulator}. Formally, we define a reward function
\[
R(\tau) = \sum_{t=1}^{T} r_t,  \quad r_t = f(\text{turn\_length}_t, \text{feedback}_t),
\]
where $\text{feedback}_t$ encapsulates sentiment and user ratings at turn $t$. This reward guides the policy $\pi(\mathbf{s}_t|\mathbf{h}_t)$, mapping the user state $\mathbf{s}_t$ (interaction history) to profile updates and response strategies that maximize cumulative engagement.

\paragraph{Algorithmic Choice: Proximal Policy Optimization} 
We adopt Proximal Policy Optimization (PPO)~\cite{Schulman2017PPO} due to its stability and sample efficiency in high‑dimensional, continuous action spaces. PPO updates policy parameters via clipped objective functions, ensuring gradual behavioral shifts—essential for maintaining user trust during personalization. The actor‑critic architecture computes advantage estimates using a value network baseline, reducing variance in gradient estimates.

\paragraph{System Integration and Related Approaches} 
Our two‑stage framework builds on recent work combining persona simulation and RL. Wang \emph{et al.}~\cite{Wang2024UserSimulator} demonstrate efficacy of a supervised‑to‑RL pipeline for realistic user simulators, while Li \emph{et al.}~\cite{Li2016DeepRL} show that policy gradient methods improve dialogue coherence and engagement.

\subsection{Prompt Manager}

Based on the trained model, we will \textbf{develop personalization strategies}:
\begin{enumerate}
\item \textbf{Dynamic Adaptation}: \textbf{Adjust the chatbot's language complexity and response style} based on the user's inferred profile.
\item \textbf{Topic Selection}: \textbf{Use the Dialogue Diversity (DD) metric} to guide topic selection and maintain engaging conversations.
\end{enumerate}
The 	\textbf{Prompt Manager} serves as the central component responsible for curating and managing conversation prompts. It receives input from the AI-Driven User Profiler to enrich the prompt based on the user's profile and current context. The Prompt Manager consists of a 	extbf{Prompt Curator} submodule that refines and adjusts prompts to suit the user's preferences and conversational style.

The user profile, constructed from the metrics in our JSON schemas, influences several key prompt parameters:

\begin{itemize}
    \item \textbf{Language Complexity Level}: Derived from the \texttt{language\_complexity} metrics, this parameter adjusts the vocabulary richness, sentence structure complexity, and technical depth of the chatbot's responses.
    
    \item \textbf{Response Detail Level}: Based on engagement patterns and response times, this determines whether responses should be concise summaries or comprehensive explanations.
    
    \item \textbf{Domain-Specific Knowledge Calibration}: Using grammar metrics and typing speed as proxies for expertise, this parameter tunes the level of domain-specific terminology and concepts included in responses.
    
    \item \textbf{Conversation Style}: Sentiment analysis and feedback patterns inform the tone and formality of responses, ranging from professional to conversational.
\end{itemize}

The 	\textbf{Dialogue Recommendation Module} generates recommendations for the next conversational steps based on user inputs and current conversation dynamics. It uses insights from the User Profiler and Interaction Tracking Engine to decide which follow-up questions or suggestions would best suit the user's current state. This module is essential for maintaining user engagement, offering proactive and personalized suggestions that align with user interests, and facilitating smooth transitions between topics.

\subsubsection{LLM/API/RAG System and Overall Flow}

The \textbf{LLM/API/RAG System} forms the core response generation mechanism for HumAIne-Chatbot, which includes large language models (LLMs), APIs, and retrieval-augmented generation (RAG) mechanisms. The enriched prompts generated by the Prompt Manager are forwarded to this system to generate highly contextual and informative responses. Depending on the user's query, it can access external knowledge via APIs or utilize vector databases to retrieve relevant information. Importantly, our approach is independent of the underlying LLM technology that supports HumAIne-Chatbot, making it adaptable to different LLMs or retrieval mechanisms as required. By designing the system to be independent of specific LLM technologies, HumAIne-Chatbot remains flexible and capable of leveraging ongoing advancements in language models, retrieval methods, and other supporting technologies. This flexibility ensures that as newer, more advanced LLMs or retrieval systems are developed, they can be seamlessly integrated to enhance the chatbot's capabilities without requiring fundamental changes to the system architecture.

\section{Evaluation Strategy}

To validate \emph{HumAIne‐chatbot}, we adopted a virtual persona–based evaluation approach inspired by VirtualXAI framework \cite{makridis2025virtualxai} that has used the approached proposed by \cite{moon2024virtual}. Instead of human recruitment, we generated 50 diverse virtual personas with systematically varied demographics (18–65+, balanced across education and occupations), communication preferences (language complexity, detail level, style), expertise levels (beginner–expert across finance, health, education, technology), behavioral patterns (patience, engagement, multitasking), and emotional profiles (stress, confidence, response speed). This design ensures scalability, reproducibility, diversity without recruitment bias, and ethical efficiency (privacy, consent, fatigue).

We performed a controlled A/B experiment contrasting two systems: (i) a \textbf{Personalized HumAIne-chatbot} using an AI-driven profiler with online adaptation; and (ii) a \textbf{Baseline Non-Personalized Chatbot} identical in all components but without adaptive profiling. Each persona interacted with both systems over 10 topic domains (e.g., career development, personal finance, health, technology, environmental sustainability). The study produced \textbf{150 sessions} (3 per persona), averaging \textbf{23 messages} and \textbf{4.13 minutes} per session (SD = 0.16), with a \textbf{100\% completion rate} and response efficiency of \textbf{334.55 messages/min}.

The \textbf{metrics framework} focused primarily on \emph{user satisfaction} (1–5 Likert, positive feedback ratio, engagement ratings, helpfulness). Complementary metrics assessed \emph{session efficiency} (duration, turns to completion, task completion) and \emph{personalization effectiveness} (response relevance, detail-level match, language complexity adaptation, style consistency). Statistical analysis employed descriptive stats, \emph{t-tests} (and Mann–Whitney U), \emph{ANOVA} for multi-factor comparisons, and \emph{effect sizes} (Cohen’s $d$). Post-hoc power analysis confirmed adequate power to detect medium–large effects.

\section{Results}

The persona cohort exhibited balanced demographics: ages 18–25 (14\%), 26–35 (16\%), 36–45 (20\%), 46–55 (16\%), 56–65 (24\%), 65+ (10\%); education levels: High School (20\%), Some College (18\%), Bachelor’s (8\%), Master’s (26\%), PhD (18\%), Professional Certification (10\%). The diversity index was \textbf{0.13}. Conversations showed \textbf{100\% completion}, \textbf{23.0} messages/session on average, \textbf{4.13} minutes/session (SD = 0.16), and throughput of \textbf{334.55} messages/min.

\begin{figure}[H]
\centering
\begin{tikzpicture}
\begin{axis}[
    ybar,
    bar width=6pt,
    width=\linewidth,
    height=0.65\linewidth,
    ymin=0, ymax=15,
    ylabel={Number of Sessions},
    xlabel={Satisfaction Score Range},
    title={Distribution of Satisfaction Scores Across All Sessions},
    xtick=data,
    xticklabels={0.05--0.10, 0.10--0.15, 0.15--0.20, 0.20--0.25, 0.25--0.30, 0.30--0.35},
    x tick label style={rotate=40, anchor=east, font=\footnotesize},
    tick label style={font=\footnotesize},
    label style={font=\footnotesize},
    title style={font=\footnotesize},
    grid=major,
    grid style={dashed,gray!30},
    enlarge x limits=0.1
]
\addplot coordinates {(1,5) (2,20) (3,20) (4,5) (5,0) (6,0)};
\end{axis}
\end{tikzpicture}
\caption{Distribution of satisfaction scores across all 50 conversation sessions.}
\label{fig:satisfaction_distribution}
\end{figure}
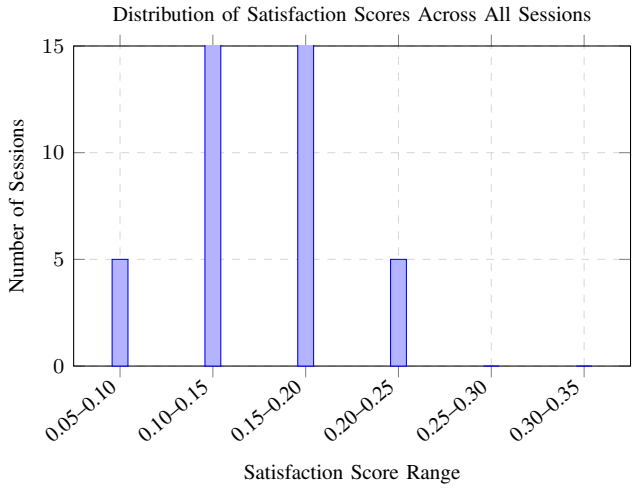

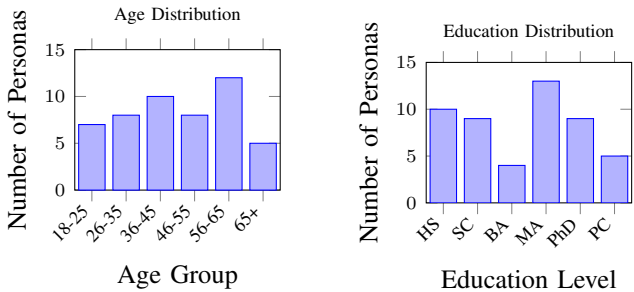
\begin{figure}[H]
\centering
\begin{subfigure}{0.48\linewidth}
\centering
\begin{tikzpicture}
\begin{axis}[
    ybar,
    width=\linewidth,
    height=0.8\linewidth,
    ylabel={Number of Personas},
    xlabel={Age Group},
    title={Age Distribution},
    ymin=0, ymax=15,
    xtick=data,
    xticklabels={18-25, 26-35, 36-45, 46-55, 56-65, 65+},
    x tick label style={rotate=45, anchor=east, font=\scriptsize},
    tick label style={font=\scriptsize},
    title style={font=\scriptsize}
]
\addplot coordinates {(1,7) (2,8) (3,10) (4,8) (5,12) (6,5)};
\end{axis}
\end{tikzpicture}
\end{subfigure}\hfill
\begin{subfigure}{0.48\linewidth}
\centering
\begin{tikzpicture}
\begin{axis}[
    ybar,
    width=\linewidth,
    height=0.8\linewidth,
    ylabel={Number of Personas},
    xlabel={Education Level},
    title={Education Distribution},
    ymin=0, ymax=15,
    xtick=data,
    xticklabels={HS, SC, BA, MA, PhD, PC},
    x tick label style={rotate=45, anchor=east, font=\scriptsize},
    tick label style={font=\scriptsize},
    title style={font=\scriptsize}
]
\addplot coordinates {(1,10) (2,9) (3,4) (4,13) (5,9) (6,5)};
\end{axis}
\end{tikzpicture}
\end{subfigure}
\caption{Demographic distribution of virtual personas: (a) Age groups, (b) Education levels.}
\label{fig:demographics}
\end{figure}

Satisfaction was the central metric. The overall mean satisfaction was \textbf{0.173} (median = 0.182, SD = 0.071). The \textbf{experimental (personalized)} condition outperformed \textbf{control} by \textbf{+45.0\%} (0.173 vs. 0.119), significant at $p < 0.001$ (t = 4.394; Mann–Whitney U = 1791.0, $p < 0.001$).

\begin{table}[H]
\centering
\caption{Descriptive Statistics for Control and Experimental Groups}
\label{tab:descriptive_stats}
\begin{tabular}{@{}lcccc@{}}
\toprule
\textbf{Group} & \textbf{Mean} & \textbf{Std Dev} & \textbf{Median} & \textbf{Range} \\
\midrule
Control (Non-Personalized) & 0.119 & 0.050 & 0.121 & 0.034--0.246 \\
Experimental (Personalized) & 0.173 & 0.071 & 0.183 & 0.056--0.318 \\
\midrule
\textbf{Difference} & \textbf{+0.054} & \textbf{+0.021} & \textbf{+0.062} & \textbf{+0.072} \\
\bottomrule
\end{tabular}
\end{table}

\begin{table}[H]
\centering
\caption{Primary Outcome Results}
\label{tab:primary_results}
\begin{tabular}{@{}lccc@{}}
\toprule
\textbf{Outcome} & \textbf{Control} & \textbf{Experimental} & \textbf{Improvement} \\
\midrule
Mean Satisfaction & 0.119 & 0.173 & +45.0\% \\
95\% CI & [0.105, 0.133] & [0.153, 0.193] & \\
\bottomrule
\end{tabular}
\end{table}

\begin{table}[H]
\centering
\caption{Secondary Outcome Measures}
\label{tab:secondary_results}
\begin{tabular}{@{}lccc@{}}
\toprule
\textbf{Metric} & \textbf{Control} & \textbf{Experimental} & \textbf{Difference} \\
\midrule
Relevance Score & 0.152 & 0.198 & +0.046 \\
Personalization Score & 0.089 & 0.234 & +0.145 \\
Expertise Alignment & 0.167 & 0.189 & +0.022 \\
Style Match & 0.134 & 0.187 & +0.053 \\
Task Achievement & 0.142 & 0.198 & +0.056 \\
\bottomrule
\end{tabular}
\end{table}

Across topic domains, improvements were consistent, with the largest gains in \emph{professional networking} (+50.3\%) and \emph{creative projects} (+50.9\%).

\begin{table}[H]
\centering
\caption{Subgroup Analysis by Topic Domain}
\label{tab:subgroup_analysis}
\begin{tabular}{@{}lccc@{}}
\toprule
\textbf{Topic} & \textbf{Control} & \textbf{Experimental} & \textbf{Improvement} \\
\midrule
Professional Networking & 0.156 & 0.235 & +50.3\% \\
Creative Projects & 0.127 & 0.192 & +50.9\% \\
Career Development & 0.083 & 0.124 & +48.8\% \\
Education and Learning & 0.100 & 0.149 & +48.5\% \\
Environmental Sustainability & 0.135 & 0.199 & +48.1\% \\
Personal Finance & 0.138 & 0.200 & +44.4\% \\
Technology Trends & 0.094 & 0.134 & +42.8\% \\
Health and Wellness & 0.095 & 0.133 & +39.5\% \\
Work-Life Balance & 0.126 & 0.176 & +39.5\% \\
Travel and Culture & 0.134 & 0.184 & +36.8\% \\
\midrule
\textbf{Overall} & \textbf{0.119} & \textbf{0.173} & \textbf{+45.0\%} \\
\bottomrule
\end{tabular}
\end{table}

We also observed cross-session learning: profile accuracy improved over \textbf{2–3 sessions}, with clearer gains in detail-level matching and language adaptation; variability persisted in complex multi-domain scenarios.

The evaluation methodology offered several important strengths. It provided systematic control and ensured reproducibility, while supporting multi-metric coverage and statistical rigor. The framework was also highly scalable and carried notable ethical advantages, as no human data collection was required. At the same time, certain limitations must be acknowledged. The synthetic nature of virtual personas cannot fully capture the complexity of real human behavior, and the structured scenarios used in the study may not reflect the nuances of natural conversation. Moreover, the evaluation was limited to short-term sessions, leaving open questions about how the system would generalize to long-horizon, free-form interactions.

\section{Conclusions}

The evaluation of \emph{HumAIne-chatbot} demonstrated the value of the \emph{VirtualXAI} approach as a scalable, reproducible, and ethically responsible methodology for assessing personalization in conversational systems. By leveraging virtual personas, the framework enables systematic exploration of user diversity while avoiding the practical and ethical constraints of human-subject studies. Looking ahead, future work will focus on enriching profiling with multimodal signals, refining reinforcement learning objectives to capture long-term user goals, and conducting pilot deployments to validate the system’s effectiveness in real-world settings.

\section{Acknowledgements}
The research leading to the results presented in this paper has received funding from the Europeans Union's funded Project HumAIne under grant agreement no 101120218.





\bibliographystyle{IEEEtran}
\bibliography{example}

\begin{thebibliography}{10}
\providecommand{\url}[1]{#1}
\csname url@samestyle\endcsname
\providecommand{\newblock}{\relax}
\providecommand{\bibinfo}[2]{#2}
\providecommand{\BIBentrySTDinterwordspacing}{\spaceskip=0pt\relax}
\providecommand{\BIBentryALTinterwordstretchfactor}{4}
\providecommand{\BIBentryALTinterwordspacing}{\spaceskip=\fontdimen2\font plus
\BIBentryALTinterwordstretchfactor\fontdimen3\font minus \fontdimen4\font\relax}
\providecommand{\BIBforeignlanguage}[2]{{%
\expandafter\ifx\csname l@#1\endcsname\relax
\typeout{** WARNING: IEEEtran.bst: No hyphenation pattern has been}%
\typeout{** loaded for the language `#1'. Using the pattern for}%
\typeout{** the default language instead.}%
\else
\language=\csname l@#1\endcsname
\fi
#2}}
\providecommand{\BIBdecl}{\relax}
\BIBdecl

\bibitem{xu2021industry}
X.~Xu, Y.~Lu, B.~Vogel-Heuser, and L.~Wang, ``Industry 4.0 and industry 5.0—inception, conception and perception,'' \emph{Journal of manufacturing systems}, vol.~61, pp. 530--535, 2021.

\bibitem{makridis2020predictive}
G.~Makridis, D.~Kyriazis, and S.~Plitsos, ``Predictive maintenance leveraging machine learning for time-series forecasting in the maritime industry,'' in \emph{2020 IEEE 23rd international conference on intelligent transportation systems (ITSC)}.\hskip 1em plus 0.5em minus 0.4em\relax IEEE, 2020, pp. 1--8.

\bibitem{makridis2023deep}
G.~Makridis, P.~Mavrepis, and D.~Kyriazis, ``A deep learning approach using natural language processing and time-series forecasting towards enhanced food safety,'' \emph{Machine Learning}, vol. 112, no.~4, pp. 1287--1313, 2023.

\bibitem{kotios2022personalized}
D.~Kotios, G.~Makridis, S.~Walser, D.~Kyriazis, and V.~Monferrino, ``Personalized finance management for smes,'' in \emph{Big Data and Artificial Intelligence in Digital Finance}.\hskip 1em plus 0.5em minus 0.4em\relax Springer, 2022, pp. 215--232.

\bibitem{brown2020language}
T.~B. Brown, B.~Mann, N.~Ryder \emph{et~al.}, ``Language models are few-shot learners,'' \emph{arXiv preprint arXiv:2005.14165}, 2020.

\bibitem{liu2023implicit}
J.~Liu \emph{et~al.}, ``Adaptive chatbot personalization through implicit user signals,'' \emph{IEEE Transactions on Affective Computing}, 2023.

\bibitem{radziwill2017evaluating}
N.~M. Radziwill and M.~C. Benton, ``Evaluating quality of chatbots and intelligent conversational agents,'' \emph{arXiv preprint arXiv:1704.04579}, 2017.

\bibitem{io2017chatbots}
H.~Io and C.~Lee, ``Chatbots and conversational agents: A bibliometric analysis,'' in \emph{2017 IEEE International Conference on Industrial Engineering and Engineering Management (IEEM)}.\hskip 1em plus 0.5em minus 0.4em\relax IEEE, 2017, pp. 215--219.

\bibitem{xu2022adaptive}
A.~Xu \emph{et~al.}, ``User-adaptive chatbot framework for enhancing customer service,'' in \emph{Proceedings of the International Conference on Intelligent User Interfaces}, 2022.

\bibitem{CercasCurry2020}
A.~Cercas~Curry and V.~Rieser, ``Conversational ux design: A practitioner's guide to the natural conversation framework,'' \emph{Journal of Interaction Studies}, vol.~21, no.~1, pp. 4--28, 2020.

\bibitem{Deriu2021}
J.~Deriu, A.~Rodrigo, C.~Gunasekara, and et~al., ``Spot the bot: A robust and efficient framework for the evaluation of conversational dialogue systems,'' in \emph{Proceedings of the 2021 Conference of the North American Chapter of the Association for Computational Linguistics: Human Language Technologies}.\hskip 1em plus 0.5em minus 0.4em\relax Association for Computational Linguistics, 2021, pp. 3815--3830.

\bibitem{Fadhil2018}
A.~Fadhil, ``Beyond patient monitoring: Conversational agents role in telemedicine \& healthcare support for home-living elderly individuals,'' \emph{ArXiv Preprint arXiv:1803.06000}, 2018.

\bibitem{moon2024virtual}
S.~Moon, M.~Abdulhai, M.~Kang, J.~Suh, W.~Soedarmadji, E.~K. Behar, and D.~M. Chan, ``Virtual personas for language models via an anthology of backstories,'' \emph{arXiv preprint arXiv:2407.06576}, 2024.

\bibitem{weizenbaum1966eliza}
J.~Weizenbaum, ``Eliza---a computer program for the study of natural language communication between man and machine,'' \emph{Communications of the ACM}, vol.~9, no.~1, pp. 36--45, 1966.

\bibitem{jasper23report}
\BIBentryALTinterwordspacing
JasperAI, ``The ai in business trend report,'' 2023, accessed:May 26, 2023. [Online]. Available: \url{https://www.jasper.ai/blog/ai-business-trend-report}
\BIBentrySTDinterwordspacing

\bibitem{sallam2023chatgpt}
M.~Sallam, ``Chatgpt utility in healthcare education, research, and practice: systematic review on the promising perspectives and valid concerns,'' in \emph{Healthcare}, vol.~11, no.~6.\hskip 1em plus 0.5em minus 0.4em\relax MDPI, 2023, p. 887.

\bibitem{makridis2024fairylandai}
G.~Makridis, A.~Oikonomou, and V.~Koukos, ``Fairylandai: Personalized fairy tales utilizing chatgpt and dalle-3,'' \emph{arXiv preprint arXiv:2407.09467}, 2024.

\bibitem{fatouros2023transforming}
G.~Fatouros, J.~Soldatos, K.~Kouroumali, G.~Makridis, and D.~Kyriazis, ``Transforming sentiment analysis in the financial domain with chatgpt,'' \emph{Machine Learning with Applications}, vol.~14, p. 100508, 2023.

\bibitem{fatouros2024can}
G.~Fatouros, K.~Metaxas, J.~Soldatos, and D.~Kyriazis, ``Can large language models beat wall street? unveiling the potential of ai in stock selection,'' \emph{arXiv preprint arXiv:2401.03737}, 2024.

\bibitem{mavrepis2024xai}
P.~Mavrepis, G.~Makridis, G.~Fatouros, V.~Koukos, M.~M. Separdani, and D.~Kyriazis, ``Xai for all: Can large language models simplify explainable ai?'' \emph{arXiv preprint arXiv:2401.13110}, 2024.

\bibitem{brandtzaeg2017people}
P.~B. Brandtzaeg and A.~F{\o}lstad, ``Why people use chatbots,'' in \emph{Internet Science: 4th International Conference, INSCI 2017, Thessaloniki, Greece, November 22-24, 2017, Proceedings 4}.\hskip 1em plus 0.5em minus 0.4em\relax Springer, 2017, pp. 377--392.

\bibitem{serban2018survey}
I.~V. Serban \emph{et~al.}, ``A survey of available corpora for building data-driven dialogue systems,'' \emph{arXiv preprint arXiv:1803.04823}, 2018.

\bibitem{li2016deep}
J.~Li, W.~Monroe \emph{et~al.}, ``Deep reinforcement learning for dialogue generation,'' in \emph{Proceedings of the 54th Annual Meeting of the Association for Computational Linguistics (Volume 1: Long Papers)}, 2016.

\bibitem{venkatesh2018evaluating}
A.~Venkatesh, C.~Khatri, A.~Ram, F.~Guo, R.~Gabriel, A.~Nagar, R.~Prasad, M.~Cheng, B.~Hedayatnia, A.~Metallinou \emph{et~al.}, ``On evaluating and comparing open domain dialog systems,'' \emph{arXiv preprint arXiv:1801.03625}, 2018.

\bibitem{fitzpatrick2017woebot}
K.~K. Fitzpatrick \emph{et~al.}, ``Delivering cognitive behavior therapy to young adults with symptoms of depression and anxiety using a fully automated conversational agent (woebot): a randomized controlled trial,'' \emph{JMIR mental health}, vol.~4, no.~2, p. e19, 2017.

\bibitem{romero2023approach}
F.~C. Romero, S.~P. Le{\'o}n, and L.~Wong, ``Approach for personalized recommendations to enhance customer service process in peruvian restaurants using openai contextual chatbot,'' in \emph{2023 IEEE XXX International Conference on Electronics, Electrical Engineering and Computing (INTERCON)}.\hskip 1em plus 0.5em minus 0.4em\relax IEEE, 2023, pp. 1--8.

\bibitem{lappromrattana2023quick}
T.~Lappromrattana and P.~Sooraksa, ``Quick prototyping of companion bots for elderly people,'' \emph{Sens. Mater}, vol.~35, pp. 1487--1495, 2023.

\bibitem{jamil2023systematic}
M.~B.~A. Jamil and D.~Shahzadi, ``A systematic review a conversational interface agent for the export business acceleration,'' \emph{Lahore Garrison University Research Journal of Computer Science and Information Technology}, vol.~7, no.~2, pp. 37--49, 2023.

\bibitem{alghiffari2023antecedents}
A.~P. Alghiffari and I.~O. Matusin, ``Antecedents of customer loyalty on ai chatbot users in banking applications,'' \emph{Jurnal Pendidikan Tambusai}, vol.~7, no.~2, pp. 18\,915--18\,927, 2023.

\bibitem{Liu2012}
B.~Liu, \emph{Sentiment Analysis and Opinion Mining}.\hskip 1em plus 0.5em minus 0.4em\relax Morgan \& Claypool Publishers, 2012.

\bibitem{han2021designing}
X.~Han, M.~Zhou, M.~J. Turner, and T.~Yeh, ``Designing effective interview chatbots: Automatic chatbot profiling and design suggestion generation for chatbot debugging,'' in \emph{Proceedings of the 2021 CHI Conference on Human Factors in Computing Systems}, 2021, pp. 1--15.

\bibitem{Moon2024VirtualPersonas}
S.~Moon, M.~Abdulhai, M.~Kang, J.~Suh, W.~Soedarmadji, E.~K. Behar, and D.~M. Chan, ``Virtual personas for language models via an anthology of backstories,'' \emph{arXiv preprint arXiv:2407.06576}, 2024.

\bibitem{Makridis2024VirtualXAI}
G.~Makridis, V.~Koukos, G.~Fatouros, D.~Kotios, M.~M. Separdani, D.~Kyriazis, and J.~Soldatos, ``Virtualxai: A user-centric framework for explainability assessment leveraging gpt-generated personas,'' in \emph{arXiv preprint arXiv:2401.13110}, 2024.

\bibitem{Li2016DeepRL}
J.~Li, W.-t. Monroe, A.~Shi, S.~Jean, A.~Ritter, and D.~Jurafsky, ``Deep reinforcement learning for dialogue generation,'' in \emph{Proceedings of the 54th Annual Meeting of the Association \ for Computational Linguistics (Volume 1: Long Papers)}, 2016.

\bibitem{Wang2024UserSimulator}
X.~Wang, Y.~Zhao, and Z.~Chen, ``Two-stage reinforcement learning for user simulator with implicit profiles in conversational ai,'' \emph{Journal of Artificial Intelligence Research}, 2024.

\bibitem{Schulman2017PPO}
J.~Schulman, F.~Wolski, P.~Dhariwal, A.~Radford, and O.~Klimov, ``Proximal policy optimization algorithms,'' \emph{arXiv preprint arXiv:1707.06347}, 2017.

\bibitem{makridis2025virtualxai}
G.~Makridis, V.~Koukos, G.~Fatouros, D.~Kotios, M.~M. Separdani, D.~Kyriazis, and J.~Soldatos, ``Virtualxai: A user-centric framework for explainability assessment leveraging gpt-generated personas,'' in \emph{2025 21st International Conference on Distributed Computing in Smart Systems and the Internet of Things (DCOSS-IoT)}.\hskip 1em plus 0.5em minus 0.4em\relax IEEE, 2025, pp. 1034--1041.

\end{thebibliography}

\end{document}